\documentclass[a4paper,10pt]{article}
\usepackage[utf8]{inputenc}
\usepackage[T1]{fontenc}
\usepackage[french,english]{babel}
\usepackage{iflang}
\usepackage{graphicx}
\usepackage{titlesec}
\usepackage{url}
\usepackage{hyperref}
\usepackage[autolanguage]{numprint}   % Composition valeur-unité
\usepackage{color}
\usepackage[usenames, dvipsnames]{xcolor}
\usepackage{tikz}
	\usetikzlibrary{patterns,arrows,decorations.pathmorphing}

\titleformat*{\section}{\fontsize{11}{13}\bfseries}
\titleformat*{\subsection}{\fontsize{10}{12}\bfseries}
\titleformat*{\subsubsection}{\fontsize{10}{12}}
\titleformat*{\paragraph}{\large\bfseries}
\titleformat*{\subparagraph}{\large\bfseries}

\usepackage[font={it}]{caption}
\makeatletter
\renewenvironment{figure}[1][\fps@figure]{
  \edef\@tempa{\noexpand\@float{figure}[#1]} 
  \@tempa\centering
}{
  \end@float
}
\makeatother
\makeatletter

\usepackage{titling}
\pretitle{\title{\IfLanguageName{english}{\enT \\ {\itshape\frT}}{\frT \\ {\itshape \enT}}}
\begin{center}
\includegraphics[width=17cm]{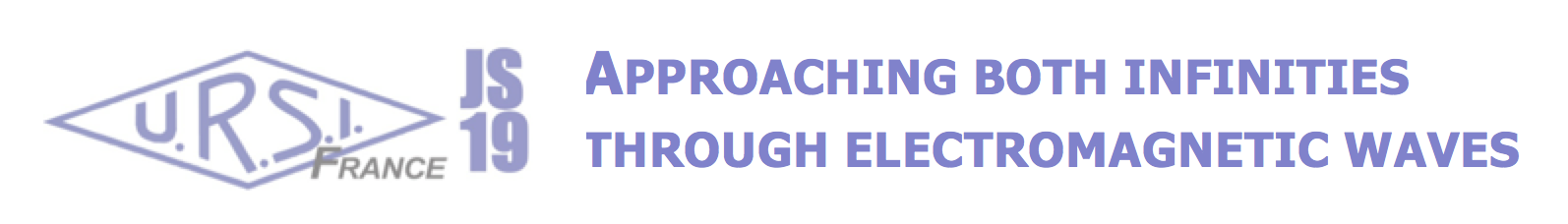}
\par\vskip 1.5em
\fontsize{14}{16}\bfseries}
\posttitle{\par\end{center}\vskip 1em\hrule\vskip 0.5em}
\preauthor{}
\postauthor{\par\vskip 0.6em\hrule}
\date{}

\newcommand{\englishtitle}[1]{\renewcommand{\enT}{#1}}
\newcommand{\enT}{}
\newcommand{\frenchtitle}[1]{\renewcommand{\frT}{#1}}
\newcommand{\frT}{}
\newcommand{\englishabstract}[1]{\renewcommand{\enA}{#1}}
\newcommand{\enA}{}
\newcommand{\frenchabstract}[1]{\renewcommand{\frA}{#1}}
\newcommand{\frA}{}
\newcommand{\englishkeywords}[1]{\renewcommand{\enK}{#1}}
\newcommand{\enK}{}
\newcommand{\frenchkeywords}[1]{\renewcommand{\frK}{#1}}
\newcommand{\frK}{}

\renewcommand{\maketitlehookd}{%
\vskip -1.5em%
\IfLanguageName{english}{%
 {\bfseries Keywords: }\enK\\{\bfseries Mots-clefs: }\frK}{%
 {\bfseries Mots-clefs: }\frK\\{\bfseries Keywords: }\enK}%
\vskip 0.5em\hrule\vskip 0.5em%
\IfLanguageName{english}{%
 {\bfseries Abstract:}\par{\small\enA}\par\vskip 1em{\bfseries R\'esum\'e:}\par{\small\frA}}{%
 {\bfseries R\'esum\'e:}\par{\small\frA}\par\vskip 1em{\bfseries Abstract:}{\small\enA}}\vskip 0.3em}

\usepackage[blocks]{authblk}

\usepackage{etoolbox}
\patchcmd{\thebibliography}{\section*}{\section}{}{}

\usepackage{array}
\usepackage{amsmath}
\usepackage{amssymb}
\usepackage{mathdots}
\usepackage{subfig}
\usepackage[load-configurations = abbreviations,alsoload=hep]{siunitx}
\DeclareSIUnit{\sample}{S}

\englishtitle{Radio detection of atmospheric air showers of particles}
\frenchtitle{La radio-détection des gerbes de particules atmosphériques}

\author[1]{A. Escudie\/}
\author[1,2]{D. Charrier\/}
\author[1,2]{R. Dallier\/}
\author[1]{D. Garc\'{\i}a-Fern\'{a}ndez\/}
\author[3]{A. Lecacheux\/}
\author[1,2]{L. Martin\/}
\author[1,2]{B. Revenu\/}

\affil[1]{SUBATECH, Institut Mines-Telecom Atlantique -- CNRS/IN2P3 -- Universit\'e de Nantes, Nantes, France}
\affil[2]{Unit\'e Scientifique de Nan\c cay, Observatoire de Paris, CNRS, PSL, UO/OSUC, Nan\c cay, France}
\affil[3]{CNRS-Observatoire de Paris, Meudon, France}

\englishkeywords{Cosmic rays, air showers, radio detection}
\frenchkeywords{Rayons cosmiques, gerbes atmosphériques, détection radio}

\englishabstract{
Since 2002, the CODALEMA experiment located within the Nan\c cay Radio Observatory studies the ultra-high energy cosmic rays (above $10^{17}$~eV) arriving in the Earth atmosphere. These cosmic rays interact with the component of the atmosphere, inducing an extensive air shower (EAS) composed mainly of charged particles (electrons and positrons). During the development of the shower in the atmosphere, these charged particles in movement generate a fast electric field transient (a few nanoseconds to a few tens of ns), detected at ground by CODALEMA with dedicated radio antennas over a wide frequency band (between 20~MHz and 200~MHz). The study of this electric field emitted during the shower development aims to determine the characteristics of the primary cosmic ray which has induced the particle shower: its nature, its arrival direction and its energy. After some theoretical considerations and a short description of the SELFAS simulation code, we will present the CODALEMA experiment, its performances and main results. At last, we will show how the EAS radio-detection technique could be used to observe very high energy gamma rays sources, with the NenuFAR radio telescope.
}
\frenchabstract{
Depuis 2002, l'exp\'erience CODALEMA, bas\'ee sur le site de l'Observatoire de radio-astronomie de Nan\c cay, \'etudie les rayons cosmiques d'ultra haute \'energie (RCUHE, au del\`a de $10^{17}$~eV) qui arrivent dans l'atmosph\`ere terrestre. Ces rayons cosmiques interagissent avec les atomes de l'atmosph\`ere, engendrant une cascade de particules secondaires charg\'ees (électrons et positrons), commun\'ement appel\'ee gerbe de particules. Lors du d\'eveloppement de la gerbe dans l'atmosph\`ere, ces particules charg\'ees en mouvement engendrent notamment l'\'emission d'une impulsion de champ \'electrique tr\`es br\`eve (de quelques nanosecondes \`a quelques dizaines de ns), que CODALEMA d\'etecte au sol avec des antennes radio d\'edi\'ees, sur une large bande de fr\'equences (entre 20~MHz et 200~MHz). L'\'etude de ce champ \'electrique \'emis lors du d\'eveloppement de la gerbe a pour but de d\'eterminer les caract\'eristiques du rayon cosmique primaire ayant engendr\'e la gerbe de particules : sa nature, sa direction d'arriv\'ee et son \'energie. Apr\`es quelques consid\'erations th\'eoriques et une courte description du code de simulation SELFAS, nous pr\'esenterons l'exp\'erience CODALEMA, ses performances et ses principaux r\'esultats. Enfin, nous montrerons comment la technique de radio-d\'etection des gerbes atmosph\'eriques pourrait \^etre  utilis\'ee pour l'observation de sources de rayons gamma d'ultra-haute \'energie.
}

\begin{document}
% !!!!!!!!!!!!!!!!!!
%% choose language
\selectlanguage{french}
\selectlanguage{english}
\maketitle

\section{Introduction}
Cosmic rays are commonly defined as relativistic atomic nuclei travelling through the Universe, and possibly arriving on Earth. The distribution of their flux according to their energy is one of the most remarkable and questioning in the modern physics, spreading over 32 orders of magnitude in flux and more than 13 orders of magnitude in energy (from a few hundred MeV up to more than $10^{20}$~eV for the most energetic ones observed so far, and that are not explained by any identified physical process). According to their energy, they constitute, just like electromagnetic radiation, a single source of information on phenomena of galactic (up to $\simeq 10^{15}$~eV) and extragalactic origin (above $10^{15}$~eV). However, the important interactions with the galactic and extragalactic magnetic fields complicate the interpretation of the collected data in order to determine their source and their nature. 

Ultra-high energy cosmic ray (UHECR) physics refers to the experimental study of cosmic rays above 10$^{14}$~eV. Above this limit, direct measurements of cosmic rays with balloons or satellites are limited by their small collecting area regarding the extremely low flux of particles (down to 1 per km$^2$ per century above 10$^{20}$~eV). Instead, one has to detect the cascade of secondary particles, called extensive air shower (EAS), which follows the interaction of the cosmic rays with the Earth's atmosphere and spread at ground over very large areas. The interaction of a cosmic ray with a nucleus of the atmosphere can occur at energies much higher than that which one can reach in the laboratory (up to 800 times higher than the 14 TeV in the center of mass in proton-proton collisions in the LHC at CERN), making them also specially interesting to refine our view on the standard model of particle physics. The primary particle characteristics (energy, arrival direction, nature) are then indirectly inferred from the EAS measurements: detected ground particles, shower front arrival time at the detector locations and the longitudinal profile of the shower which is the distribution of the number of particle in the shower over time along the shower axis. The experimental challenge stands in the ability to determine these 3 quantities with a sufficient level of accuracy \cite{2016288}. EAS are commonly observed by ground-based particle detectors and by telescopes that observe the fluorescent light emitted by the atmospheric di-nitrogen molecules excited by the passage of the cascade. The Pierre Auger observatory in Argentina \cite{auger}, the largest cosmic ray observatory in the world today, combines these two detection techniques. However, a third detection method exists: the radio detection of EAS. First developed during the 1960's, then abandoned due to technological limitations but fully revisited at the digital era since 2002, this technique is based on the fact that, during the shower development in the atmosphere, the charged particles in movement generate a coherent, transient electric field. The latter has two main origins: transverse current variation induced by the geomagnetic field, and the charge excess mechanism~\cite{Kahn206,Askaryan,2009APh....31..192A,2015APh....69...50B,radioemissionprd}. Since 2002, the CODALEMA experiment located within the Nan\c cay Radio Observatory is one of the pioneering and promotor experiment of this revival of the radio detection of EAS, which is today adopted by a flurry of experiments in the world. Coupled with an increasingly sophisticated understanding of the processes involved allowing the use of high-performance simulation codes, EAS radio-detection now reaches a level of maturity sufficient to match the more traditional methods for fundamental properties of UHECR.

In the most common way, the observations are carried out in the restricted range \SI{[20-80]}{\mega\hertz} (noted MF in the following, for Medium Frequencies) by experiments such as AERA~\cite{ThePierreAuger:2015rma}, Tunka-Rex \cite{BEZYAZEEKOV201589}, TREND \cite{2012arXiv1204.1599O}, Yakutsk experiment~\cite{Knurenko:2016dck} or LOFAR~\cite{2013AA...556A...2V}. The use of this band is mainly due to man-made broadcasting at low and medium frequencies (AM, FM bands) leading to the choice of relative low sampling rates (\SI{\sim 200}{\mega\sample\per\second}) of the digitizers used by experiments such as AERA and LOFAR. However, CODALEMA~\cite{dallier:hal-01164680,BenoitICRC2017} works with a sampling rate of \SI{1}{\giga\sample\per\second}, making it possible to extend observations up to the band  \SI{[110-200]}{\mega\hertz}. The main limitation of the frequency band is then due to the bandwidth of the antenna used, which is \SI{[20-200]}{\mega\hertz} for CODALEMA, referred to as Extended Medium Frequencies (EMF) in the following.

CODALEMA is today the supporting experiment of the EXTASIS experiment \cite{EXTASIS}, an array of 7 low-frequency antennas, which takes advantage of its existing infrastructure. EXTASIS aims to reinvestigate the [1-10] MHz band, and especially to study the so-called "Sudden Death" contribution, the expected electric field emitted by shower front when hitting the ground level. Currently, EXTASIS has confirmed some results obtained by the pioneering experiments, and tends to bring explanations to the other ones, for instance the role of the underlying atmospheric electric field \cite{EXTASIS}.

In this paper, we will first briefly describe the processes involved in the generation of the EAS. Then, we will highlight the mechanisms at the origin of the transient electric field, and describe its complete expression as it is implemented in the SELFAS simulation code that we developed. In a third section, we will present the CODALEMA experiment and its principle of functioning, whose main results and performances will be exposed in the next section. At last, we will show how the EAS radio-detection technique could be used to observe very high energy gamma rays sources, with the NenuFAR radio telescope located nearby CODALEMA.

\section{Extensive Air Showers}

\begin{figure}[h!]
\begin{center}
\subfloat[]{
  \includegraphics[width=0.36\textwidth]{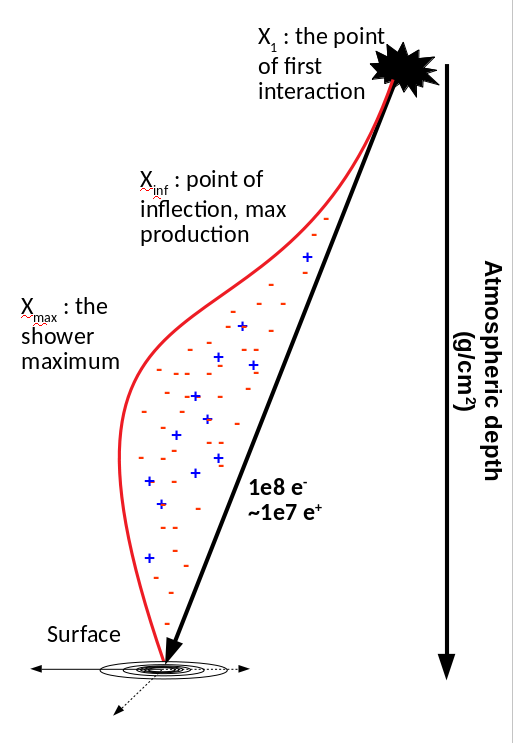}
}
\subfloat[]{
  \includegraphics[width=0.4\textwidth]{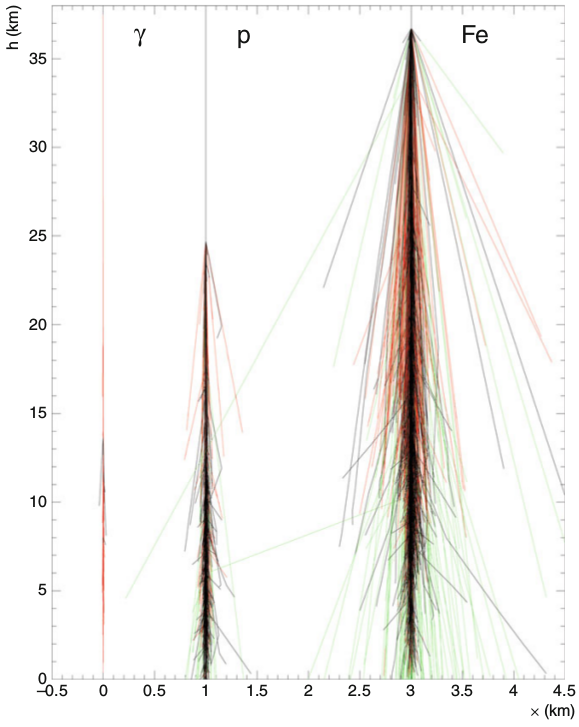}
}
\caption{(a): Sketch of the geometry of an extensive air shower, with the parameters $X_1$, $X_{inf}$ and $X_{max}$. The approximate number of electrons and positrons is indicated for a primary proton with an energy of \SI{10^{17}}{\electronvolt}. (b): Extensive shower development for different primary cosmic ray. The three primaries are simulated with an energy of \SI{10^5}{\giga\electronvolt}. Red component is for electromagnetic, green component for muons and black for hadrons. Adapted from \cite{Grieder:1339647}.}
\label{gerbe}
\end{center}
\end{figure}

Extensive air showers are initiated by a series of interactions of high energetic primary cosmic ray hadron with the atmosphere constituent ($N_2$, $O_2$ and $Ar$). The secondary charged particles propagate in the atmosphere, forming a hadron cascade longitudinally along the primary cosmic ray trajectory, with a lateral dispersion due to their transverse momentum. During the interactions of the primary particle with the nuclei of the atmosphere, some pions $\pi^0$ and $\pi^\pm$ are created. The particles $\pi^0$ decay in two $\gamma$ which produce $e^-$ and $e^+$ by pair production. The $e^+$ are rapidly absorbed in the medium while the $e^-$ could create $\gamma$ by Bremsstrahlung. These $\gamma$ decay in a $e^+$--$e^-$ pair. This is the electromagnetic component of the shower, containing \SI{90}{\%} of the primary particle energy \cite{Huege:2016veh}, that can reach very large dimensions depending on the primary cosmic ray energy. A simple illustration of extensive air shower is shown in figure \ref{gerbe}.(a), with the point of first interaction $X_1$, the point of inflection $X_{inf}$ where the production of secondary particles in the shower is maximum and the depth of maximum development $X_{max}$ until which the number of particle increases. From this point, the available energy in the shower front is insufficient and the number of particles absorbed in the atmosphere is larger than the number of particles produced: the number of particles in the shower decreases till the shower extinction.
The value $X$ is calculated by integrating the density of air (taking its changes into account) along the trajectory of the shower, from the point of entry of the air shower at the top of the atmosphere to the point in question (units in \si{\gram\per\centi\metre\squared}, increasing up to \SI{\sim1000}{\gram\per\centi\metre\squared} at sea level ). The longitudinal profile refers to the number of particles along the development of the shower. 

The shower development depends on the primary cosmic ray characteristics: its energy, its arrival direction and its nature. For very energetic primary, the shower front could reach the ground, and the maximum shower development occurs near the sea level. Inversely, a low energetic primary induces a small shower with a maximum of development reached at higher altitude in the atmosphere and only muons and neutrinos can reach the ground. Depending on the nature of the primary, the point of first interaction $X_1$ is different, because the cross-section depends on the mass number $A$ and on the energy $E$ (see Fig. \ref{gerbe}). The lateral dispersion is also very different for the three primaries shown here (gamma photon, proton or iron nucleus), though they have the same energy of \SI{10^5}{\giga\electronvolt}. On Fig. \ref{gerbe}, only the secondary particles with an energy above \SI{10}{\giga\electronvolt} are presented. If we considerer all the secondary particles, the $\gamma$-induced air shower has a dispersion radius -- defined as the lateral dispersion measured from the shower axis at sea level -- of $\simeq{}$\SI{100}{\metre}, the proton-induced air shower has a dispersion radius of $\simeq{}$\SI{500}{\metre} and the iron-induced air shower of \SI{\sim1000}{\metre}. The shower maximum $X_{max}$ is also different for the three primaries, being sensitive to both the primary mass and the primary energy. Knowing the latter, moreover rather easily measurable, measuring the $X_{max}$ gives hints on the primary cosmic ray nature. $X_{max}$ is directly observed in fluorescence telescopes (though with a quite low duty cycle of about 14~\% \cite{ABRAHAM2010227}), while for particle detectors and radio it is most of the time reconstructed thanks to shower simulations by Monte-Carlo associated to particle interaction codes.

\section{Radio emission from EAS}
\subsection{Physics of radio transient emission mechanisms}
During the shower development in the atmosphere, the shower composed of charged particles in movement generates a coherent radio emission which has two origins: the charge excess mechanism, and the transverse current variation induced by the geomagnetic field~\cite{Kahn206,Askaryan,2009APh....31..192A,2015APh....69...50B,radioemissionprd}. The latter is the main mechanism of radio emission. Secondary electrons and positrons in the air shower are accelerated and multi-scattered by the geomagnetic field, and separated thanks to the Lorentz force $\overrightarrow{F}=q\overrightarrow{v}\times\overrightarrow{B}$ (with $q$ the particle charge,  $\overrightarrow{v}$ its velocity and $\overrightarrow{B}$ the geomagnetic field vector). The resulting currents will be, on average, perpendicular to the shower axis and are called "transverse currents". They vary with the air shower development and the number of secondary particles which first grows, reaches a maximum, and then decreases as the shower dies out. This time variation will lead to electromagnetic radiation \cite{Huege:2016veh}. The resulting electric field is linearly polarized and aligned along $\overrightarrow{F}$ (thus along $\overrightarrow{n}\times\overrightarrow{B}$), where the direction of propagation of the secondary particles can be assimilated with the shower axis $\overrightarrow{n}$. The intensity of the electric field depends on the arrival direction of the primary cosmic ray, in particular on the angle $\alpha$ between the arrival direction of the shower and the geomagnetic field.\\

At the second order, the charge excess mechanism comes in addition to the geomagnetic contribution. Actually, at high energy, the processes of pair production and bremsstrahlung dominate. As the shower develops, the average energy decreases and other processes appear, as Compton scattering, delta-ray production, knock-on electrons, and positron annihilation with pre-existing free atmospheric electrons, resulting in a net negative charge excess of $\approx$ 10 - 20 $\%$. The ionization electrons are contained in the curved shower front, while the heavier positive ions stay behind. The same scheme as for the geomagnetic mechanism occurs: during the shower development, the net negative charge grows, reaches a maximum and finally decreases when the shower dies out. The time variation of the evolution of the charge excess also leads to radio transient emission \cite{Huege:2016veh}. Contrarily to the geomagnetic one, here the electric field vector is radially polarized with respect to the shower axis, and its orientation depends on the location of an observer. The overall resulting emission is a superposition of these two mechanisms, and appears as a fast electric field transient lasting few tens of nanoseconds, which can be detected by large bandwidth antennas and fast acquisition systems.

\subsection{Formalism and prediction of the radio emission from EAS -- The SELFAS simulation code}
In addition to the shower development simulations codes like CORSIKA \cite{CORSIKA}, CONEX \cite{Bergmann:2006yz} or AIRES \cite{AIRES}, based on particle physics hadronic interaction models such as or Sybill \cite{Ahn:2009wx} or EPOS \cite{Pierog:2009zt} or QGSJET \cite{Ostapchenko:2004ss}, several authors have implemented the calculation of the electric field emitted by the charged particles of the shower. Most of them have chosen a microscopic approach, using the simulated particle tracks as elementary sources of electric field (\cite{Huege:2013vt,ALVAREZMUNIZ2012S187}), and adding all the contributions at the observation point. Developed within the frame of CODALEMA, SELFAS (Simulation of Electric Field of Air Showers) uses the same approach, based on a complete and detailed formalism. But, instead of simulating each interaction in the shower along its development, SELFAS uses pre-defined particle distributions obtained from a large set of previous shower simulations (\cite{2009APh....31..243L}), thus considerably reducing the computing time. The complete formalism used in SELFAS is obtained by rewriting Maxwell's equations in the form of two coupled equations by using the scalar and vector potentials and by using the Lorentz gauge which can be expressed as $\nabla\cdot\textbf{A}+\mu\epsilon\frac{\partial{\Phi}}{\partial{t}}=0$:
\begin{align}
\nabla\cdot\textbf{D} = \nabla\cdot(\epsilon\textbf{E}) &= \epsilon\nabla\cdot\textbf{E} = \rho \nonumber \\ 
&= \epsilon_0\nabla(-\frac{\partial\textbf{A}}{\partial{t}}-\nabla{\Phi}) = \rho \nonumber \\ 
&= -\nabla^2\Phi - \nabla(\frac{\partial\textbf{A}}{\partial{t}}) = \frac{\rho}{\epsilon} \nonumber \\ 
&\Rightarrow \boxed{\nabla^2\Phi - \frac{1}{c_n^2}\frac{\partial^2\Phi}{\partial{t}^2} = -\frac{\rho}{\epsilon}}
\label{Bscalar}
\end{align}
where $c_n$ is the speed of light in the dielectric medium. The same calculation can be made for the vector potential giving:
\begin{align}
 \boxed{\nabla^2\textbf{A} - \frac{1}{c_n^2}\frac{\partial^2\textbf{A}}{\partial{t}^2} = -\mu_0\textbf{J}}
 \label{Bvector}
\end{align}
To solve equations \ref{Bscalar} and \ref{Bvector}, we use retarded solutions known as Green functions, giving: 
\begin{equation}
\label{Phi}
 \Phi(\textbf{x},t) = \frac{1}{4\pi\epsilon}\int\mathrm{d}^3x'\frac{1}{R}\left[\rho(\textbf{x}',t')\right]_\mathrm{ret}
\end{equation}
\begin{equation}
\label{A}
 \textbf{A}(\textbf{x},t) = \frac{\mu_0}{4\pi}\int\mathrm{d}^3x'\frac{1}{R}[\textbf{J}(\textbf{x}',t')]_\mathrm{ret}
\end{equation}
where $R=|\textbf{x}-\textbf{x}'|$ and $t'$ is the retarded time defined as $t'=t-\frac{1}{c_n}|\textbf{x}-\textbf{x}'(t')|$. Using equations \ref{Phi} and \ref{A}, we can derive the electric field and obtain:
\begin{equation}
\label{Equ}
\textbf{E}(\textbf{x},t) = \frac{1}{4\pi\epsilon}\int\mathrm{d}^3x'\left[\frac{\textbf{\^R}}{R}[\rho(\textbf{x}',t')]_\mathrm{ret}+\frac{\textbf{\^R}}{c_nR}\left[\frac{\partial\rho(\textbf{x}',t')}{\partial{t}'}\right]_\mathrm{ret}-\frac{1}{c_n^2R}\left[\frac{\partial\textbf{J}(\textbf{x}',t')}{\partial{t}'}\right]_\mathrm{ret}\right]
\end{equation}
where $\textbf{\^R}=\frac{\textbf{x}-\textbf{x}'}{|\textbf{x}-\textbf{x}'|}$. The calculation of the electric field having been derived from the Maxwell equations, we need to define the source whose the electric field is to be calculated. Let us define the instant $t_1$ before which the charge density in all space is zero. Let us consider a neutral atom. At $t_1$ and position $\textbf{x}_1$, a point-like source separates from the atom with the charge $-q$ and travels in a straight line at a constant speed until its sudden stop at instant $t_2$ and position $\textbf{x}_2$. The associated charge density, which certifies that the charge is conserved, can be written as:
\begin{eqnarray}
\rho(\textbf{x}',t') & = & -q\delta^3(\textbf{x}'-\textbf{x}_1)\Theta(t'-t_1) \nonumber \\ 
&& \: +q\delta^3(\textbf{x}'-\textbf{x}_1)-\textbf{v}(t'-t_1))[\Theta(t'-t_1)-\Theta(t'-t_2)] \\
&& \: +q\delta^3(\textbf{x}'-\textbf{x}_2)\Theta(t'-t_2) \nonumber
\label{CD}
\end{eqnarray}
where $\delta$ is the Dirac delta function and $\Theta$ the Heaviside step function. The associated current density can be written as:
\begin{equation}
 \textbf{J}(\textbf{x}',t')=q\textbf{v}\delta^3(\textbf{x}'-\textbf{x}_1)-\textbf{v}(t'-t_1))[\Theta(t'-t_1)-\Theta(t'-t_2)]
 \label{CuD}
\end{equation}
Finally, the electric field can be expressed as:
\begin{eqnarray}
\label{Equ4}
\textbf{E}(\textbf{x},t) & = & \frac{q}{4\pi\epsilon}\left\lbrace\ -\frac{\textbf{\^R}_1}{R_1^2}\Theta(t-t_1-R_1/c_n)-\frac{\textbf{\^R}_1}{c_nR_1}\delta(t-t_1-R_1/c_n) \nonumber \right.\\ 
&+& \: \left. \left[\frac{\textbf{\^R}_1}{\kappa R^2}(\Pi(t',t_1,t_2))\right]_\mathrm{ret}+\frac{1}{c_n}\frac{\partial}{\partial t}\left[\frac{\textbf{\^R}_1}{\kappa R}(\Pi(t',t_1,t_2))\right]_\mathrm{ret}-\frac{\textbf{v}}{c_n^2}\frac{\partial}{\partial t}\left[\frac{1}{\kappa R}(\Pi(t',t_1,t_2))\right]_\mathrm{ret} \nonumber \right.\\
&+& \: \left. \frac{\textbf{\^R}_2}{R_2^2}\Theta(t-t_2-R_2/c_n)+\frac{\textbf{\^R}_2}{c_nR_2}\delta(t-t_2-R_2/c_n) \right\rbrace
\end{eqnarray}
in which we have used $t'=t-R_i/c_n$ for $t'<t_1$ and $t'>t_2$. First (creation of the particle) and third (stop of the particle) lines of equation \ref{Equ4} contain a static Coulomb field which contributes at $t_1$  and $t_2$ respectively and an impulse radiation field with the form $\frac{\textbf{\^R}_i}{c_nR_i}\delta(t-t_i-R_i/c_n)$ which is due to the charge conservation and the use of a realistic charge density. The impulse radiation field is due to the changes in the charge density, which thus can be paired with the second line. The second line is equivalent to the Heaviside-Feynman expression used in \cite{2012arXiv1211.3305R} to calculate the electric field of a particle track, and is inseparable from the impulse radiation field. The equation \ref{Equ4} is able to describe the electric field created by a particle track \cite{Garcia-Fernandez:2017yss}, and has been implemented in the third version, SELFAS3, of the SELFAS Monte Carlo code.

\section{Instrumental setup}\label{sec:instrumentalSetup}

\begin{figure}[htb]
\centering
\includegraphics[width=0.7\textwidth]{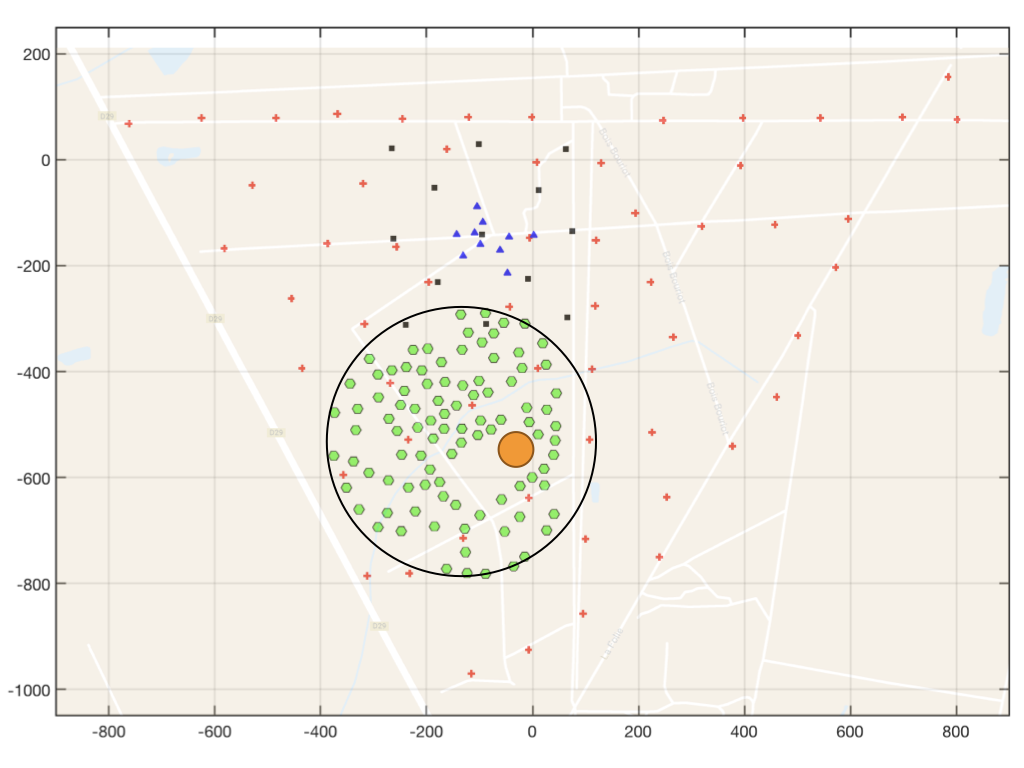}
\caption{Map of the Nan\c cay observatory (North on top), featuring some of the setups currently involved in the radio detection of EAS. Red crosses feature the 57 CODALEMA autonomous radio detection stations, black squares the 13 scintillators, blue triangles the 10 antennas of the compact array. The location of the LOFAR FR606 LBA station is figured by the orange circle, while the NenuFAR radiotelescope mini-arrays are the green hexagons which fit into the black circle. The scale is indicated on the axis. The area covered by CODALEMA is $\simeq$1.1~km$^2$.}\label{fig:arrayMap}
\end{figure}

CODALEMA~\cite{dallier:hal-01164680,BenoitICRC2017} is hosted since 2002 by the Nançay Radioastronomy Observatory. It is one of the pioneering experiments that have participated in the rebirth of radio detection of cosmic rays. Over the years, the experiment has seen the development of a large collection of detectors, intended to study the properties of the radio emission associated with cosmic ray induced air showers in the energy range from $10^{16}$ to \SI{10^{18}}{\electronvolt}. In its current version, CODALEMA consists essentially of:
\begin{itemize}
\item{a square array (0.4 $\times$ \SI{0.4}{\kilo\metre\squared}) of 13 particle scintillator counters (surface detector),}
\item{a set of 57 so-called ``autonomous'' crossed dipoles and synchronized by GPS dating, operating in the EMF band, distributed over \SI{1}{\kilo\metre\squared},}
\item{a so-called ``Compact Array'' of 10 cross-polarized antennas, arranged in a star shape of \SI{150}{\metre} extension and whose signal acquisition (in the MF band) is triggered by the particle detector.}
\end{itemize}
The autonomous station array is purely self-triggered, meaning that each station is independent. Transients coming from cosmic ray air showers or any other source (noise, planes...) are either stored on a distant disk for off-line analysis or directly sent to a central DAQ able to build on-line the event based on several station signals, respecting several selection criteria, which offers a large noise rejection factor (more than 99~\%) and a very good efficiency on cosmic ray air shower transient detection. A crosscheck can be made off-line with the events detected by the particle detector or any of the triggered instruments (Compact Array or Low-Frequency stations). CODALEMA is installed in a very rich instrumental environment (Fig.~\ref{fig:arrayMap}), since the Nan\c cay radio astronomy station houses a set of radio telescopes covering wavelengths from decametre to decimetre (frequency range [10 - 3500]~MHz) \cite{webnancay}. As many international developments aim to finalise the SKA project, a special effort is being made at the moment on the decametric domain, with the installation of the FR606 station of the international radio telescope LOFAR \cite{2013AA...556A...2V} and the construction, already well advanced, of the NenuFAR array (``New Extension in Nan\c cay Upgrading LOFAR'') \cite{NenuFAR}, which is described in section \ref{sec:gamma}.\\

\section{Event reconstruction, results and performances}

\begin{figure}[p]
\begin{center}
\subfloat[]{
  \includegraphics[width=0.5\textwidth,height=0.37\textwidth]{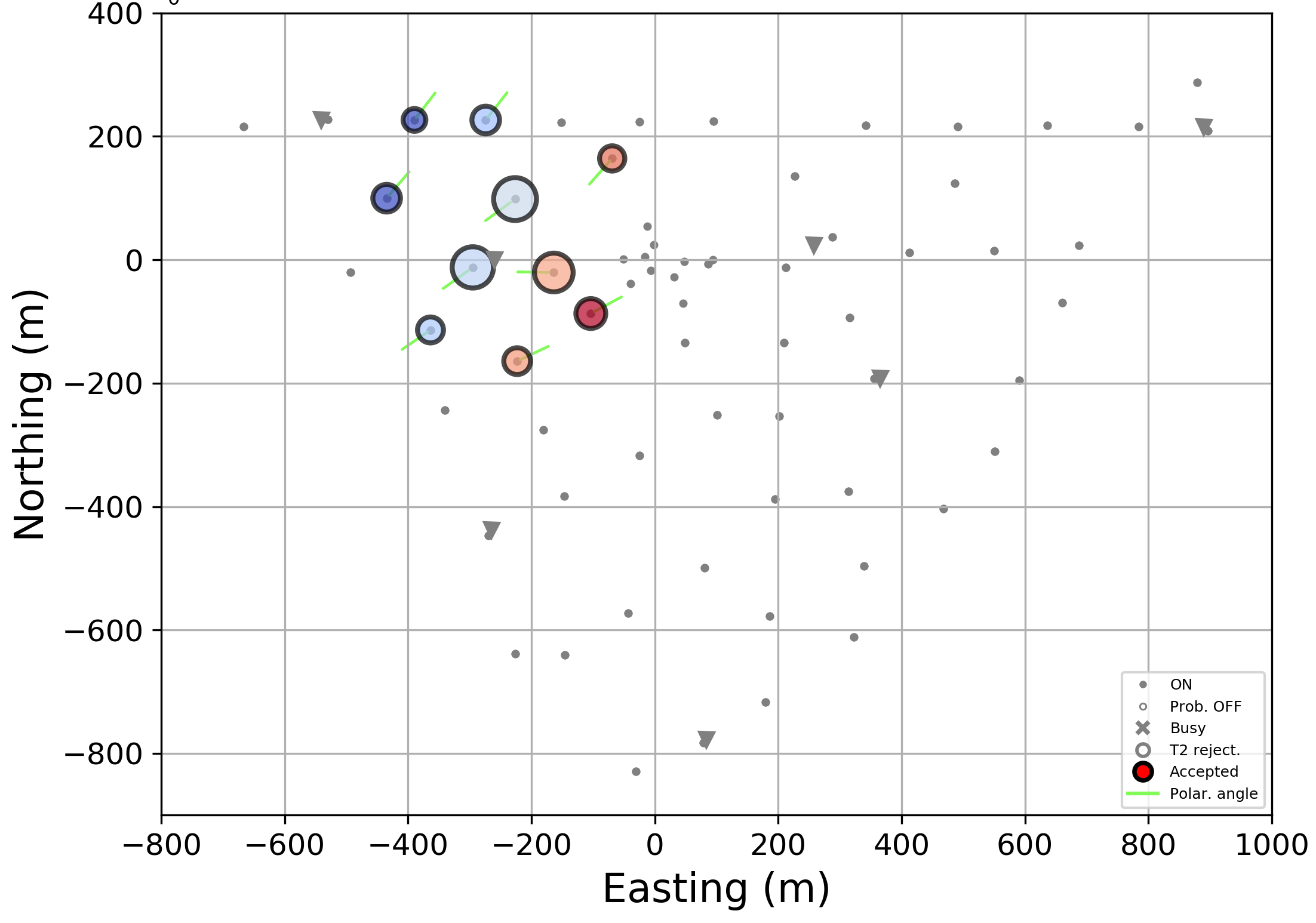}
}
\subfloat[]{
  \includegraphics[width=0.5\textwidth]{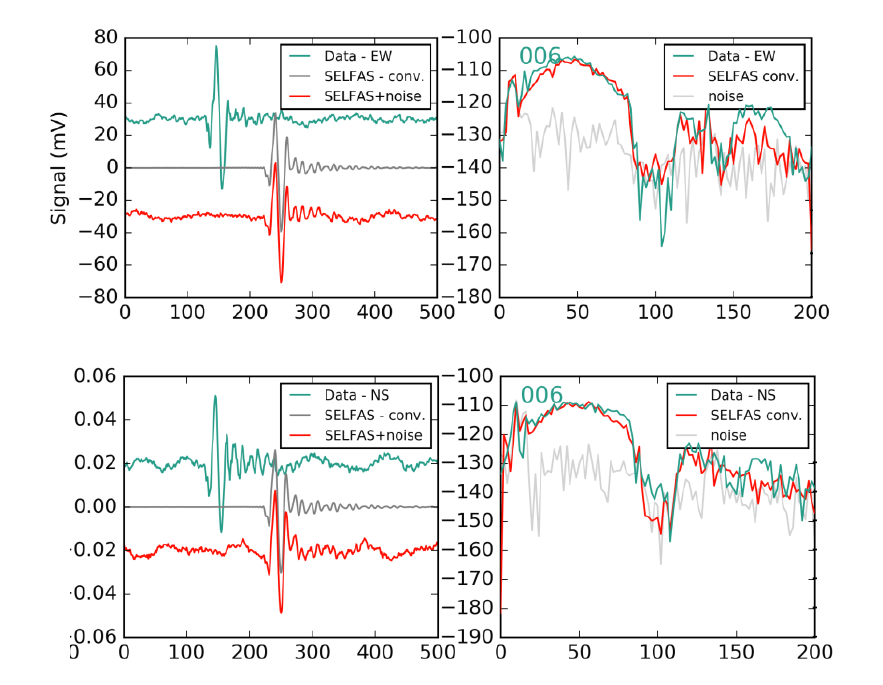}
}
\\
\subfloat[]{
  \includegraphics[width=0.48\textwidth,height=0.36\textwidth]{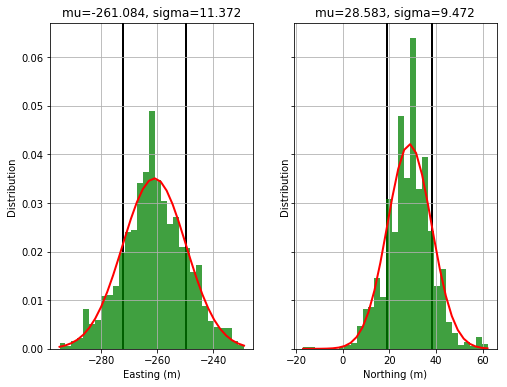} 
}
\hfill
\subfloat[]{
  \includegraphics[width=0.45\textwidth,height=0.36\textwidth]{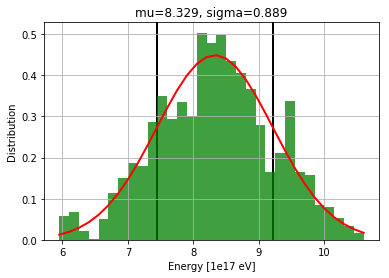}
}
\\
\subfloat[]{
    \begin{tikzpicture}
    \node[anchor=south west,inner sep=0] (image) at (0,0) {\includegraphics[width=0.45\textwidth]{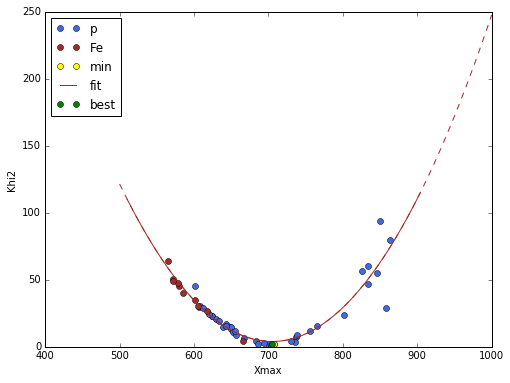}};
    \begin{scope}[x={(image.south east)},y={(image.north west)}]
        \draw[dashed] (0.535,0.1) -- (0.535,0.55);
        \draw (0.535,0.6) node[align=left]{\scriptsize $X_{max} = $\SI{705\pm19}{\gram\per\centi\metre\squared}};
    \end{scope}
        \end{tikzpicture}
}
\caption{Illustration of the reconstruction of one event. (a): map of the event. The grey dots represent the radio antennas of the experiment, the colored circles represent the triggered antennas. The size of the circles is proportional to the signal strength received by the antennas, the color gradient represent the timing order in which the signal is seen by each antenna (from blue to red). (b), left canvas: transient detected by one antenna (green) in the EW (above) and NS (below) polarization, associated to the transient simulated with SELFAS3 (red). (b), right canvas: power spectral density received by an antenna (green) in the EW (above) and NS (below) polarization, associated to the power spectrum density simulated with SELFAS3. The good agreement between data and the simulation shows that the radio detection technique and the response of our radio antennas are well mastered. The other plots (c), (d) and (e) indicate the accuracy on the shower parameters. The red curve is a Gaussian fit of the distributions, the black vertical lines represent the 1$\sigma$ confidence level. The distributions are obtained by propagating the uncertainties on the measured values of the electric field. See text for more details.}
\label{recoevt}
\end{center}
\end{figure}

For the 57 standalone antennas, no particle trigger is sent as mentioned in section \ref{sec:instrumentalSetup}. Transients detected in coincidence on several of these standalone antennas build a ``radio coincidence'', characterized by an average radio event time that will be compared to the particle detector event. The criterion is that at least three standalone antennas are triggered within a time interval compatible with the propagation of a plane wave at the speed of light. The radio event is promoted as an actual shower if its timing is compatible with the timing of the scintillators and if the reconstructed arrival directions (the direction of arrival (DoA) is reconstructed using a plane fit) agree within \SI{20}{\degree}~\cite{dallier:hal-01164680,BenoitICRC2017}. Then, for each actual shower, a set of simulations is produced using SELFAS3: 40 protons showers and 10 iron showers at an arbitrary energy (\SI{10^{17}}{\electronvolt}) with the corresponding DoA, on a virtual array. For different shower core positions, we compare the measured values of the electric field in the MF band to the simulated electric field times a scaling factor to correct from the arbitrary energy. The best agreement is obtained by minimising the chi-squared (see \cite{LilianICRC2017} for more details). A method of error propagation is used to calculate the errors on the estimation of the shower parameters. At each step of the comparison, an error on the measured values of the electric field is randomly calculated within the gaussian distribution of the electric field values. This error is added to the electric field values, and the comparison procedure is repeated. After the propagation of the errors, the estimated parameter distributions of the shower is obtained, see figure \ref{recoevt}. To fully exploit the capabilities of the CODALEMA instruments, the reconstruction takes into account the high frequency data leading to a reconstruction in the EMF band (very useful for inclined showers), as well as the information from the Compact Array leading to an hybrid reconstruction. After these improvements, at the end of the analysis chain, the obtained accuracy is \SI{15}{\metre} on the core position, \SI{20}{\%} on the primary energy and \SI{20}{\gram\per\centi\metre\squared} on the atmospheric depth of the maximum of the shower development noted $X_\mathrm{max}$. Figure~\ref{recoevt} shows the reconstruction performances on an event detected by CODALEMA. % on March, 9$^\text{th}$, 2017.

CODALEMA is a multi-wavelength experiment observing cosmic-ray induced air-showers in \SI{[20-200]}{\mega\hertz}. Its unique capabilities permit to better constrain the reconstruction of the cosmic ray properties, and to propose soon a composition of the cosmic rays with a sufficient level of accuracy. Contrarily to most of the worldwide radio experiments, the reconstruction of all of the cosmic ray parameters by CODALEMA is made only with the radio signal, which demonstrates the self-consistency and the maturity of the radio-detection method for EAS observations. However, even though the noise rejection rate reaches almost \SI{100}{\%} and even though we have a very good efficiency on cosmic ray air shower transient detection, there is still a lot of parasitic transients. As an example, for two weeks of data taking, 52.5 millions of events are built by the central DAQ and only \SI{1}{\%} are selected after the different stages of rejection, corresponding roughly to \numprint{550000} events. In these set of events, only 34 are tagged as cosmic ray events (\SI{0.006}{\%} of the $\sim$\numprint{550000} recorded events). Different methods are being studied to increase the rejection rate and to improve the efficiency of detection of cosmic rays.

\section{Potential of the radio-detection technique for the detection of very high energy gamma rays}\label{sec:gamma}
With the instrumental development of the H.E.S.S., MAGIC and VERITAS ground telescopes, the Fermi satellite and the flurry of results accompanying them, gamma astronomy is now reaching a mature stage, and this field routinely offers astronomers and physicists images and spectra from sources that are more and more numerous and distant. The big project for the future, CTA, will allow for the first time a deep study of the gamma sky observed at energies from a few tens of GeV to more than 100 TeV. However, one of the drawbacks of the technique is that the current telescopes observing the Cerenkov radiation of the showers created by the photons in the atmosphere have a rather low operating cycle, due to severe constraints on the necessary nighttime observation conditions (no Moon, no clouds ...). Moreover, if CTA should be able to reach or even exceed the energy limit currently held by HAWC ($\simeq$ 100 TeV), the standard average sensitivity of other telescopes is restricted to a few tens of TeV for the most energetic photons.\\
One of the specificities of UHECR detection is that it is impossible to know \textit{a priori} in which direction to ``look", and therefore the detection systems, whatever they are, must have maximum angular acceptance and availability in order to cover all possible directions of arrival. The UHECR radio detection arrays are no exception to this rule. Given the energy threshold of detection found for the UHECR on observations with individual antennas (of the order of few $10^{16}$~eV), this method is not suitable to detect air showers generated by hundreds of TeV gamma rays, though they are similar, in terms of charge contents, to that of UHECR. However, the signals of several antennas may be combined \textit{a priori} or \textit{a posteriori} in a given direction, in order to gain in sensitivity of detection, which also reduces the angular field of view: it is the principle of interferometers, widely used on past and current generations of so-called ``digital" radio telescopes. Our idea is therefore to combine the principle of detection of atmospheric shower electric field transients, used in CODALEMA, with the ability of antennas to be phased towards a known source direction. Unlike the case of UHECR, we then know where the signal should come from. The immediate advantage is the gain in detection sensitivity, which varies in direct proportion to the square root of the number of antennas involved. Furthermore, combining several antenna beams into one single, directive beam makes the detector much less sensitive to noises sources generally located close to or at the horizon, and which generate the high noise transient rate observed at a single antenna level. Another advantage is the possibility of reaching a useful observation cycle close to 100~\%, since the day/night alternation and the weather have no influence on the detection itself (except in case of large local atmospheric electric fields during thunderstorms). Based on our experience of ultrafast radio transient detection and on the context of the Nan\c cay radio astronomy station which hosts the NenuFAR radio telescope, we propose to explore the possibilities offered by this idea and to try for the first time to detect the radio signal produced by  ultra-high energy gamma of a few hundred TeV to a few tens PeV from an identified astrophysical source - provided they can produce gammas at such energies.\\

\begin{figure}[htb]
\centering
\includegraphics[width=0.33\textwidth]{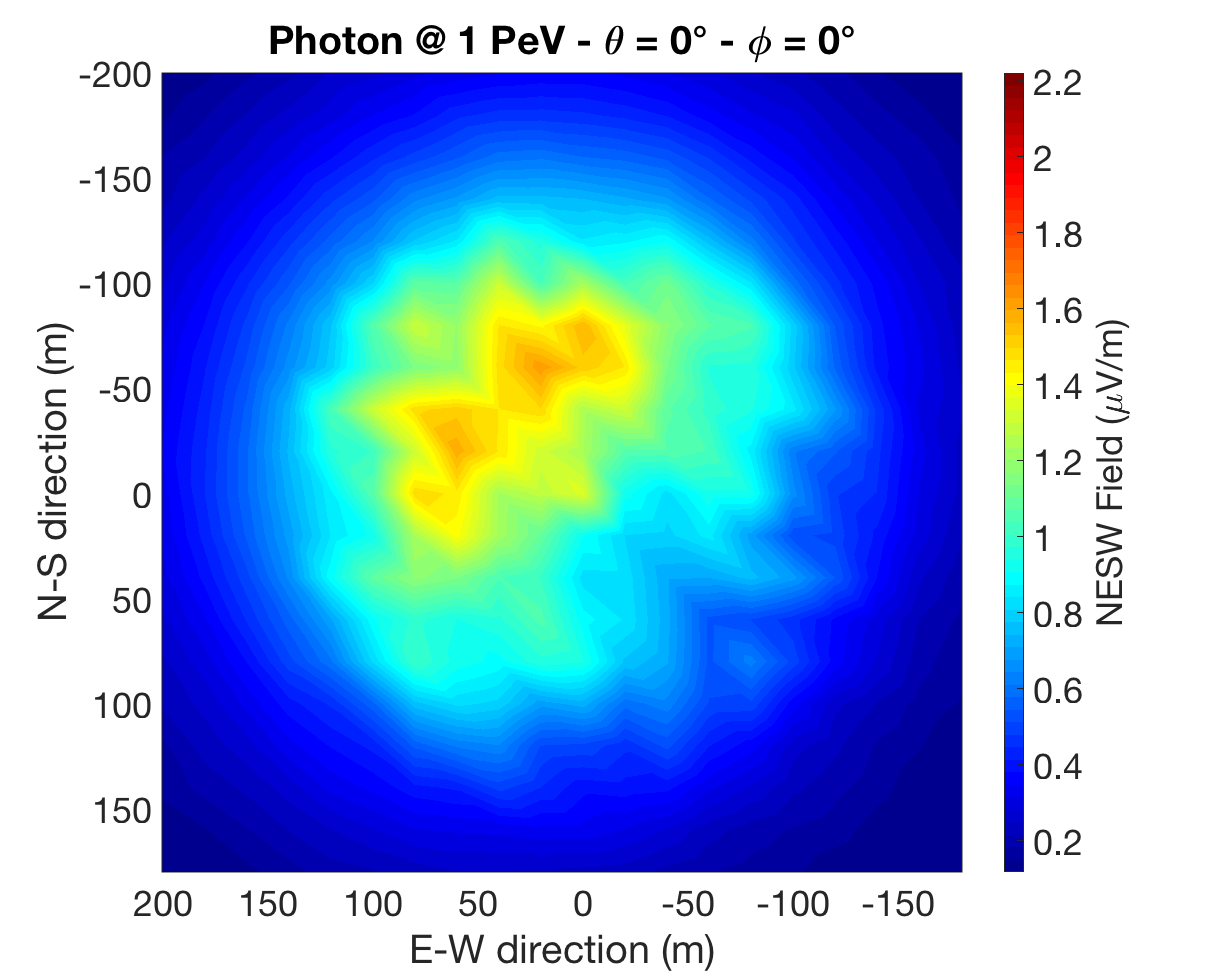}\includegraphics[width=0.33\textwidth]{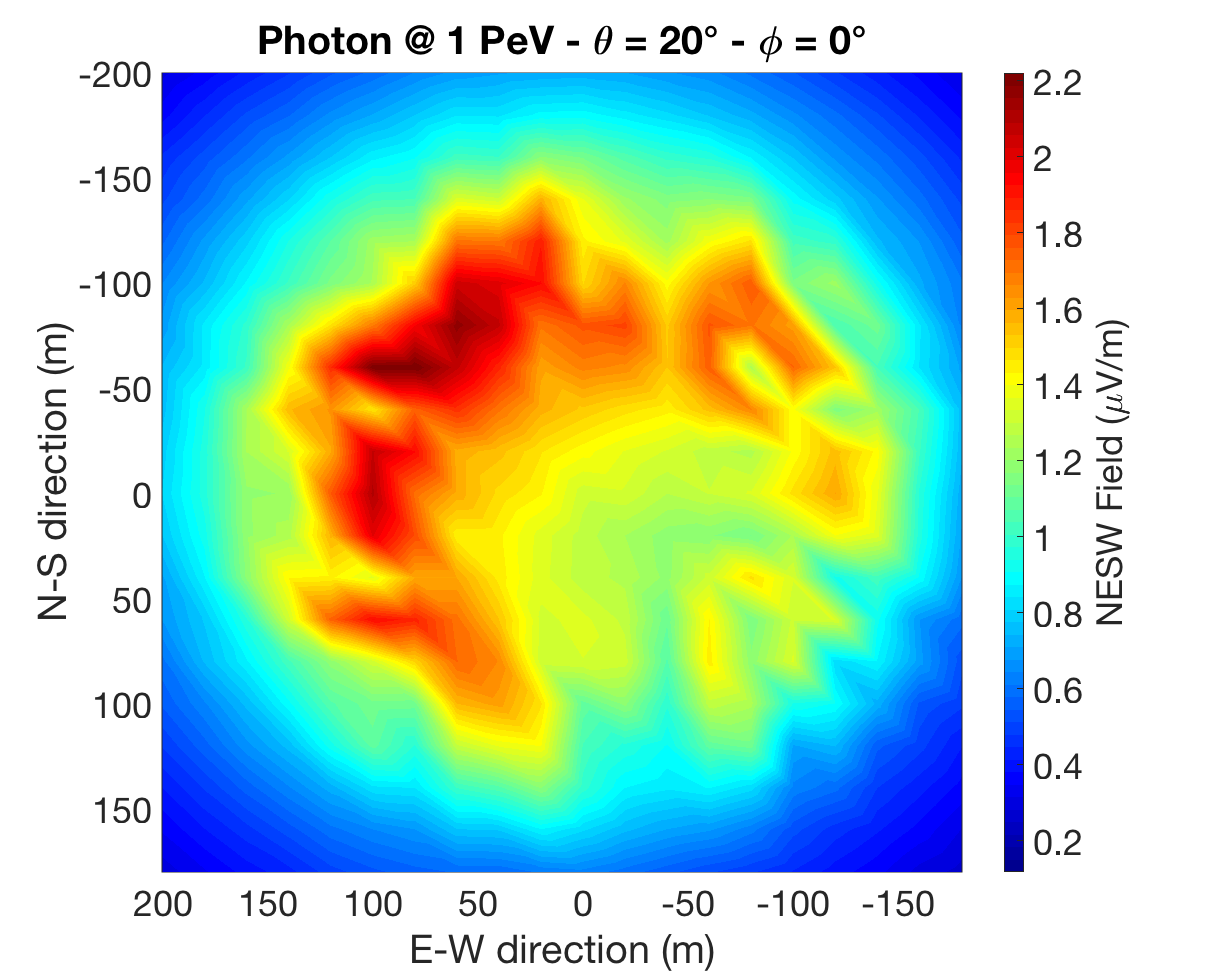}\includegraphics[width=0.33\textwidth]{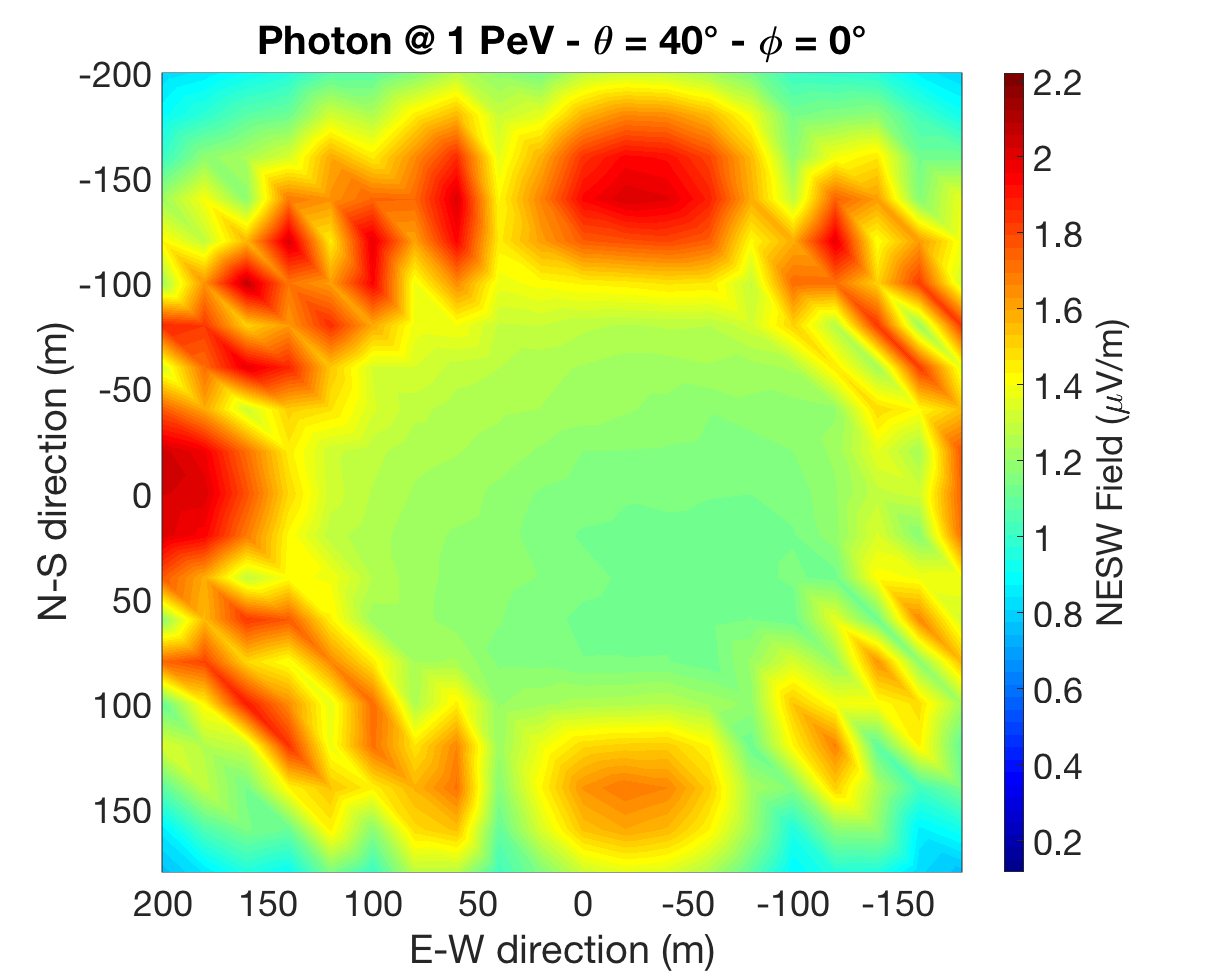}
\includegraphics[width=0.33\textwidth]{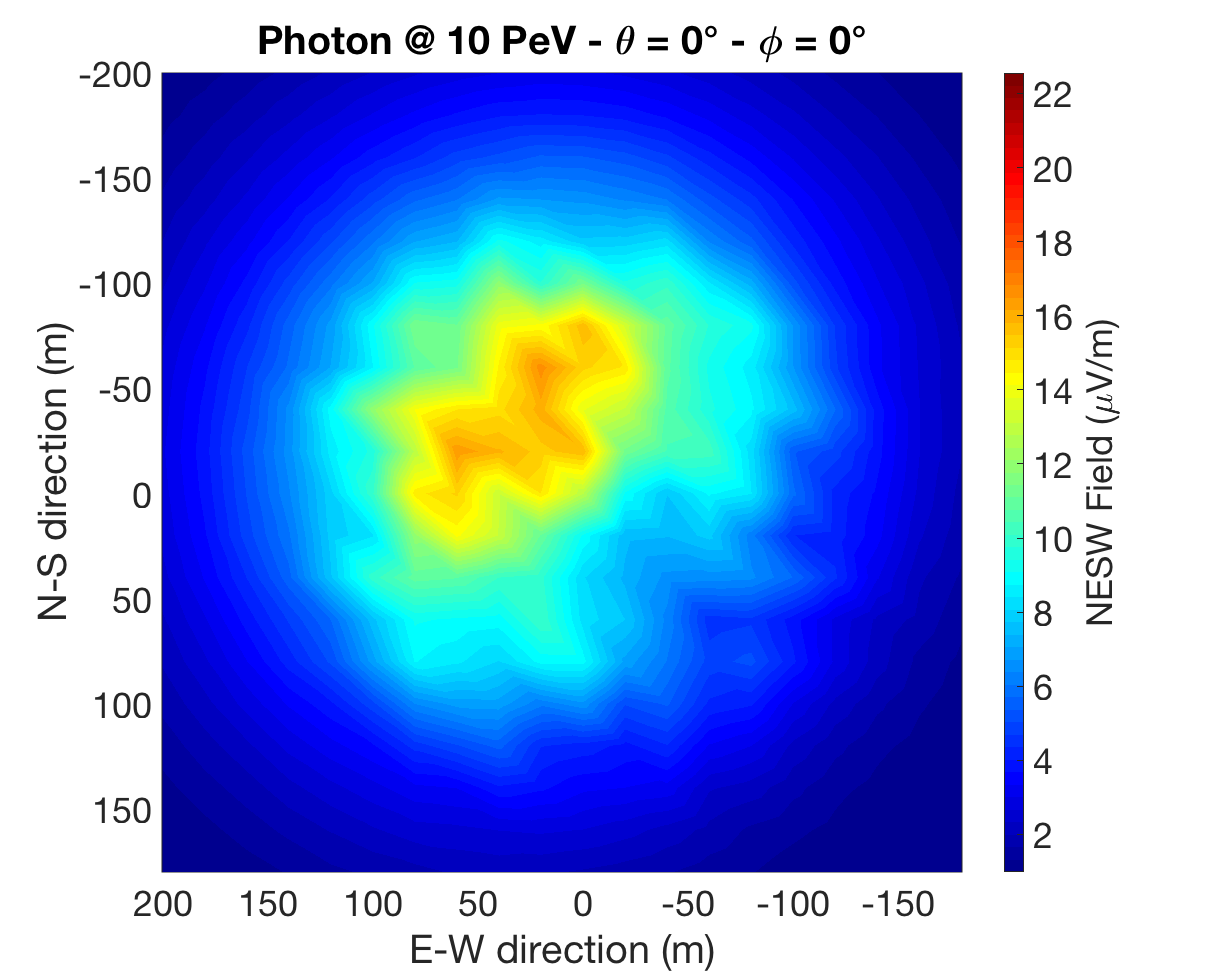}\includegraphics[width=0.33\textwidth]{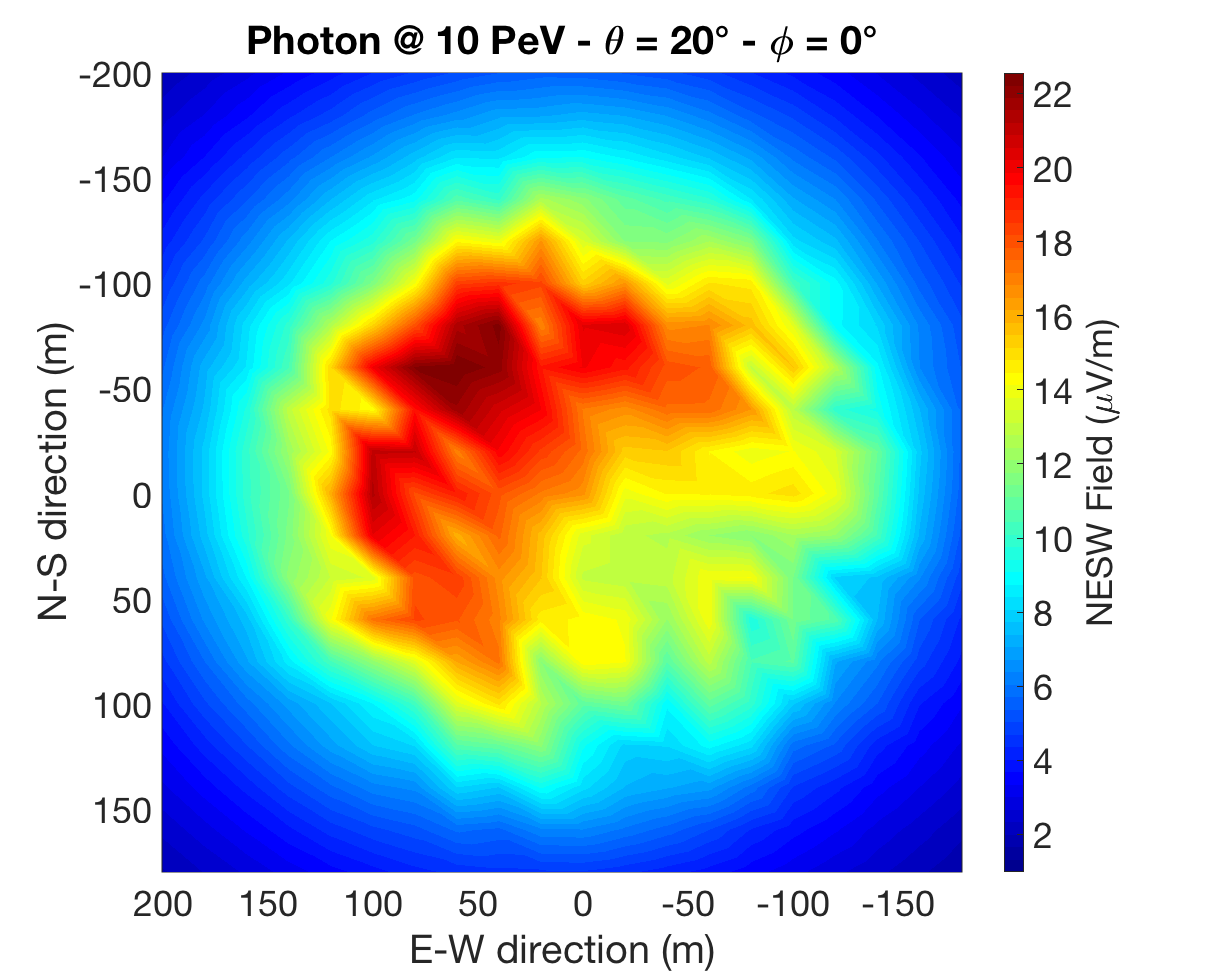}\includegraphics[width=0.33\textwidth]{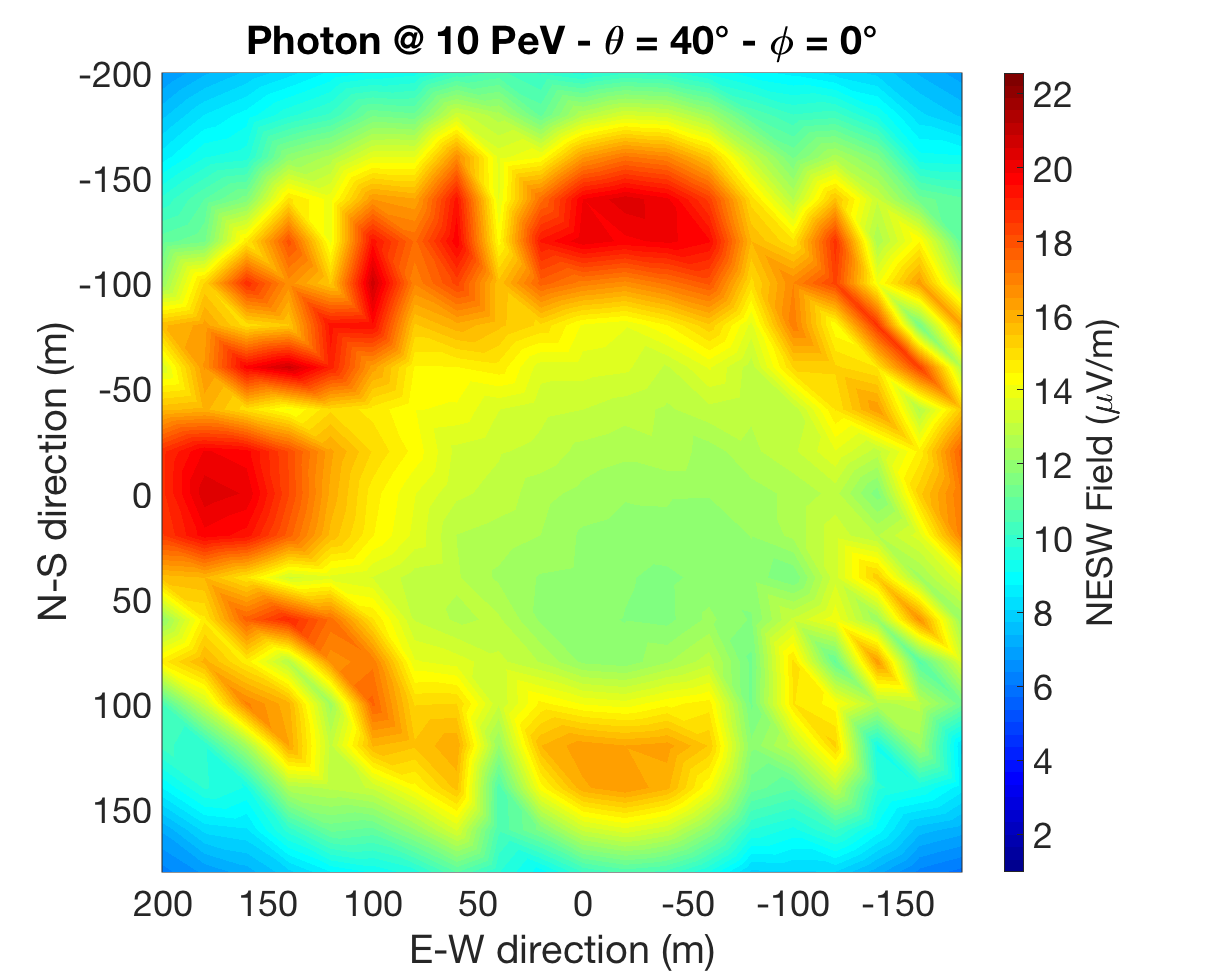}
\caption{Top figures: simulations of a 1 PeV photon shower coming from East at a zenith angle of 0, 20 and 40$^o$, in NE-SW polarization, observed in the [24-82]~MHz band. The footprint area fits well with the area of NenuFAR. The colour scale gives the electric field value in $\mu$V$.m^{-1}$. The Cerenkov ring is clearly visible with increasing zenith angle. Bottom figures: same, for 10~PeV photon showers. The colour scale is the same for all the plots.}\label{fig:pevsimulations}
\end{figure}

Also known as the LOFAR Super Station (LSS) in Nan\c cay, NenuFAR \cite{NenuFAR} is a digital radio telescope consisting - in the long term - of 1824 crossed-dipoles antennas arranged in 96 mini-arrays (hereafter MA$_{19}$) of 19 antennas\footnote{In 2018, 54 MA$_{19}$ of those 96 are installed and operational}  (see Fig.~\ref{fig:arrayMap}). It has recently been recognized as a "pathfinder" of SKA. The NenuFAR antennas use the active amplifier developed for CODALEMA and are identical to the ones of its Compact Array (see section~\ref{sec:instrumentalSetup} and \cite{dallier:hal-01164680,BenoitICRC2017}). The 96 MA$_{19}$ are distributed over an approximate circle of 500 m radius (0.2~km$^2$), see Fig.~\ref{fig:arrayMap}. In classical operation, each of the MA$_{19}$ is analogically phased in any direction of the sky and in the whole or any sub-band of the [10-85]~MHz frequency window, leading to an instantaneous sensitivity $\sqrt{19}$ times higher than that of a single antenna. The whole NenuFAR thus has an instantaneous sensitivity much higher than that reached by a single LOFAR station of 96 single antennas  \cite{2013AA...556A...2V}, like the FR606 unit present at Nan\c cay.\\
As already mentioned, the specificity of the gamma-ray sources is that we know their position in the sky, thus the arrival direction of the high-energy photon, contrarily to the UHECR case. It is therefore possible to use one or several mini-arrays phased together in the direction of the source and use them as a trigger on transient events. We can take benefit of the large memory depth of the remaining MA$_{19}$, which would be triggered by the trigger arrays, to roll-back in time and find the transient in their memories. Each MA$_{19}$ would then be a sampling point of the shower footprint at ground, exactly as it is done in UHECR radio detection arrays with single antennas. This would allow knowing the electric field distribution and use it to recover the properties of the primary photon, notably its energy thanks to simulations.\\
As an illustration, Fig. \ref{fig:pevsimulations} shows a simulation of the electric field that would be produced by a PeV and a 10 PeV gamma ray. Though sources able to produce such high energy photons (so-called ``PeVatrons'') are not yet known, but expected around the Galactic Center (unfortunately not visible from Nan\c cay), it is rather encouraging: the electric field at 10 PeV is already almost the energy detection threshold for one single antenna, and the signal strength scales linearly with the energy. With a subsequent gain in sensitivity thanks to the phasing of several mini-arrays, we should be able to detect hundreds of TeV photons showers and maybe open the door to a new detection technique of very high energy gamma rays.

\section{Conclusion}
Since 2002, the CODALEMA experiment located within the Nançay radio-astronomy observatory has been studying ultra-high energy cosmic rays arriving in the Earth's atmosphere through the study of the electric field emitted during the development of the shower of secondary charged particles. Among the precursors of this method, nowadays commonly used worldwide, CODALEMA is a laboratory of major technological innovations (autonomous triggering, hybrid reconstruction, very wide frequency band) and its high-performance antennas are exported to other sites or instruments (AERA on Auger, NenuFAR, prototypes in China, Greece$\ldots$). CODALEMA has largely contributed to demonstrate that this detection method is able to determine the characteristics of the primary cosmic ray with the only radio signal, thanks to comparisons with shower's electric field simulations performed with the SELFAS code which has been specially developed. Today, SELFAS allows a faithful and realistic reconstruction of all the characteristics of the electric field at all frequencies. Its use is essential for the identification of the primary particle that initiated the shower.\\

Besides the long-standing study of charged UHECR, we also propose today to exploit the unique environment of CODALEMA to explore with NenuFAR the possibilities of radio-detection of atmospheric showers initiated this time by very high energy gammas. The central idea here is to combine a large set of NenuFAR antennas (several tens) in the direction of known sources emitting gammas (catalogs HESS, MAGIC, VERITAS ...) to significantly increase the sensitivity of detection and use the capabilities triggering on ultra fast transients controlled within the framework of CODALEMA. This would make it possible to observe the sources with a useful cycle close to 100~$\%$, which is not possible with the current techniques of the aforementioned Cerenkov telescopes or their successors (i.e. CTA). Coming from the infinitely big, all these infinitely small, elementary particles inform us about the great structures of the Universe that emitted them. \\

\section*{Acknowledgements}
We thank the R\'egion Pays de la Loire for its financial support of the Astroparticle group of Subatech and in particular for its contribution to the EXTASIS experiment, and the PNHE (Programme National Hautes Energies) from the French institutes IN2P3 and INSU for having also always supported the CODALEMA experiment, both financially and scientifically. Lastly, we thank the technical support teams of the Nançay Radioastronomy Observatory for valuable assistance during and after the deployment of the EXTASIS antennas.

\bibliographystyle{myunsrt} % or \bibliographystyle{unsrt}
\bibliography{ursi.bib} % mon fichier de base de données s'appelle biblio.bib

\end{document}